\documentclass[aps,prd,preprintnumbers,twocolumn,groupedaddress]{revtex4}

\usepackage{graphicx}
\usepackage{hyperref}

\newcommand{\beq}{\begin{eqnarray}}
\newcommand{\eeq}{\end{eqnarray}}

\begin{document}

\title {Two flavor QCD and Confinement - II}

\author{Guido Cossu}
\email{g.cossu@sns.it}
\affiliation {Scuola Normale Superiore, Piazza dei Cavalieri 7,
  I-56127 Pisa and Dipartimento di Fisica dell'Universit\`a di Pisa and INFN, 
Sezione di Pisa, Largo Pontecorvo 3, I-56127 Pisa, Italy}
\author{Massimo D'Elia}
\email{delia@ge.infn.it}
\affiliation{Dipartimento di Fisica dell'Universit\`a di Genova and INFN,
Sezione di Genova, Via Dodecaneso 33, I-16146 Genova, Italy}
\author{Adriano Di Giacomo}
\email{digiaco@df.unipi.it}
\affiliation {Dipartimento di Fisica dell'Universit\`a di Pisa and INFN, 
Sezione di Pisa, Largo Pontecorvo 3, I-56127 Pisa, Italy}
\author{Claudio Pica}
\email{pica@bnl.gov}
\affiliation{Physics Department, Brookhaven National Laboratory,
Upton, NY 11973-5000, USA}

\begin{abstract}
This paper is part of a program of investigation of the chiral 
transition in $N_f=2$ QCD, started in Ref.~\cite{nf2-I}.
Progress is reported on the understanding of some possible systematic
errors. A direct test of first order scaling is presented.

\end{abstract}
\preprint{BNL-NT-07/30, IFUP-TH/2007-14, GEF-TH-03-07}
\maketitle

\section{Introduction}
\label{intro}

The order of the chiral transition of two-flavor QCD is not established yet, 
despite its great relevance to understand important aspects of color confinement
and of the structure of the QCD phase diagram~\cite{Review}.
A natural order parameter in that limit is the chiral condensate, even 
if empirically one finds that also deconfinement happens at the same
critical temperature $T_c$, as found by looking at the susceptibility
of the Polyakov loop or by direct studies of order parameters
constructed in the framework of specific models of color 
confinement~\cite{fullmono1,fullmono2,bari}.

A renormalization group analysis plus $\epsilon$-expansion 
can be made around $m \simeq 0$, assuming that the relevant degrees of
freedom for the chiral transition are scalar and pseudoscalar
fields~\cite{wilcz1,wilcz2,wilcz3}. The result is that
for $N_f = 2$, if the $U_A(1)$ anomaly term is still effective at 
the transition, i.e. if the  $\eta '$ mass does not vanish at $T_c$,
an infrared stable fixed point may exist in the universality class
of the three dimensional $O(4)$ spin model. Two possibilities are
therefore left open: a first order order phase transition, or a second 
order phase transition with $O(4)$ critical indexes.

Two completely different scenarios correspond to those two
possibilities. Since, contrary to the first order case, 
second order phase transitions are
unstable against the explicit breaking of the underlying symmetry,
in the second case one would have a crossover instead of a real
phase transition for small but non-zero quark masses. That would 
imply the possibility of going continuously from a confined to a 
deconfined state of matter, in contrast
with the idea of confinement being an absolute property
of strongly interacting matter at zero temperature and of
deconfinement being an order-disorder transition associated 
to a change of symmetry~\cite{hooft}.
A second consequence would be the presence of a crossover line, at finite
mass, in the region of high temperature and small baryon chemical
potential $\mu_B$ of the QCD phase diagram, thus implying a critical
point~\cite{Stephanov} to connect with the first order line which is
supposed to exist at low temperatures and large chemical
potentials.
No such point is expected to exist if the transition at $\mu_B = 0$ is first order.
No critical point has been found up to now in experiments with heavy 
ions, but the question is still open~\cite{tri1,tri2,tri3}.

The problem can in principle be solved numerically using lattice QCD
simulations, by means of a finite size
scaling (FSS) analysis leading eventually to a precise
determination of the critical indexes of the transition. 
However technical difficulties are encountered in this program.
First of all, huge computational resources are needed to make 
numerical simulations of the system which are close enough both to the 
chiral and to the thermodynamical limit (i.e. with small enough quark
masses and large enough spatial volumes).
Second, a FSS analysis is made intricate by the fact
that the system has two relevant scales: the correlation length $\xi$
of the order parameter and the inverse quark mass
$1/m_q$. In particular it is possible to write the following
FSS ansatz for the free energy density ${\cal L}/kT$ around the chiral
critical point
\beq
\frac{\cal L}{kT} \simeq L_s^{-d} \phi \left(\tau L_s^{1/\nu}, am_q L_s^{y_h} \right) \, .
\label{scal2}
\eeq
$\cal L$ depends on two different scaling variables instead of one
as is the case for simpler systems, like e.g. the quenched theory.
$L_s$ is the spatial size, $\tau$ is the reduced temperature $\tau=(1-T/T_c)$,
$\nu$ is the critical index of the correlation length ($\xi \sim
\tau^{-\nu}$) and $y_h$ is the magnetic critical index. From
Eq.~(\ref{scal2}) one can then derive the FSS for the specific heat
\beq
C_V - C_0 \simeq  L_s^{\alpha/\nu} \phi_c \left(\tau L_s^{1/\nu}, am_q
L_s^{y_h} \right) \, ,
\label{scalcal}
\eeq
where $C_0$ stems from an additive renormalization~\cite{BREZIN82},
and for the susceptibility $\chi$ of the chiral order parameter 
\beq
\chi_m -\chi_0 \simeq L_s^{\gamma/\nu} \phi_\chi \left(\tau L_s^{1/\nu}, am_q L_s^{y_h} \right) \, .
\label{scalord}
\eeq
A few groups have investigated the problem on the lattice with
staggered~\cite{fuku1,fuku2,colombia,karsch1,karsch2,jlqcd,milc}
or Wilson~\cite{cp-pacs} fermions. The common procedure has been
to use approximate versions of the scaling laws (\ref{scalcal}) and
(\ref{scalord}), usually assuming to be already in the infinite volume limit.
No clear answer has been found, but there exists a general
prejudice in favor of a second order chiral transition.

In Ref.~\cite{nf2-I} we have approached the problem by use 
of staggered fermions and a novel strategy for the
FSS analysis which, together with the availability of 
relevant resources of computer power, has allowed us to achieve 
some progress. We have decided to keep one of the scaling
variable fixed, so as to reduce the FSS analysis,
Eqs.~(\ref{scalcal}) and (\ref{scalord}), again to one
variable. In order to do that, after choosing a value 
for the critical index $y_h$ appropriate to a given universality class,
we have performed a series of runs
at variable spatial size $L_s$ and quark mass $m_q$, 
keeping the quantity $a m_q L_s^{y_h}$ fixed. The choice for 
the index $y_h$ implies an a priori assumption about the critical
behavior, which can then be carefully checked without any 
approximation looking at the residual scaling. 
In particular in Ref.~\cite{nf2-I} we have chosen 
to test the $O(4)$ critical
behavior, or better $O(2)$, which is more appropriate for the 
case of staggered fermions at non zero lattice spacing~\cite{jlqcd}:
we have therefore fixed $y_h=2.49$, which happens to be the same both  
for $O(4)$ and for $O(2)$.
Our results for the chiral susceptibility, for the specific heat and 
for the equation of state have then shown a clear inconsistency 
with both the $O(4)$ and $O(2)$ scaling hypothesis, thus giving
clear evidence against the possibility that those critical behaviors 
can describe the QCD phase transition for $N_f = 2$ in the chiral limit.
In Ref.~\cite{nf2-I} we did not perform the analogous scaling test
assuming a first order. We do that in the present paper (see Sect.~\ref{firstord}).
In Ref.~\cite{nf2-I} we have obtained some evidence in favor of a 
first order transition, by keeping the first scaling variable fixed. We 
then do the approximation of spatial
size large compared to the inverse pion mass. With this approximation 
the scaling laws read
\beq
C_V - C_0 \simeq  am_q^{-\alpha/(\nu y_h)} \phi_c \left(\tau L_s^{1/\nu} \right) \, ,
\label{scalcal2}
\eeq
\beq
\chi_m -\chi_0 \simeq am_q^{-\gamma/(\nu y_h)} \phi_\chi \left(\tau L_s^{1/\nu} \right) \, .
\label{scalord2}
\eeq
We have checked Eqs.~(\ref{scalcal2})-(\ref{scalord2}) and we found disagreement 
with $O(4)$, $O(2)$ and agreement with a weak first order.
A further result of our study was that a simple analysis of
the dependence of the pseudocritical temperature on the quark mass, 
when correctly taking into account the dependence of the
physical lattice scale on $m_q$, does not allow to
discriminate between a second order and a first order critical 
behavior.
In Ref.~\cite{nf2-I} also the magnetic equation of state was 
consistent with weak first order (see also Sect.~\ref{firstord} below).

The results obtained in Ref.~\cite{nf2-I} must be considered as
the starting point of an accurate study of the problem.
Indeed several questions have been left open, 
which deserve further analysis.
First of all, the preliminary evidence in favor of a first order
transition should be directly confirmed by a series of runs in which
$a m_q L_s^{y_h}$ is kept fixed according to first order critical
behavior, i.e. $y_h = 3$.
Second, our evidence against a second order transition in the
universality
class of $O(4)$ or $O(2)$ should be verified against all possible 
systematic effects which could have influenced our results, 
in particular: {\it i)} In Ref.~\cite{nf2-I} we have used an non-exact
$R$ algorithm~\cite{RALG}, even if being quite conservative in the
choice of the integration step for molecular dynamics; that could lead to 
systematic errors which, in principle, could influence the determination of the order
of the phase transition. {\it ii)} In Ref.~\cite{nf2-I} we have used 
a standard gauge and fermionic action, and a temporal extent $L_t =
4$, corresponding to a lattice spacing $a \sim 0.3$ fm around the phase 
transition ($T = 1/(a L_t)$). Critical behavior is a typical
infrared phenomenon. Nevertheless ultraviolet cut-off effects 
could in principle have some influence on it so that the continuum limit
of our results should be checked by using a smaller lattice spacing, i.e.
an improved action and/or a larger value of $L_t$.
Finally, if the chiral transition is really first order, however weak,
one should find signals of metastability in the physical observables
when going to large enough volumes; no convincing signals were found in
previous literature, nor in Ref.~\cite{nf2-I}, and the question
should be clarified, by exploring larger volumes.

Answering to all previous questions represents a difficult and 
computationally demanding program, which we partially carry out
in the present paper. In particular we address the question
related to the use of a non-exact algorithm in Section~\ref{rhmc},
where some of the results obtained with the R algorithm in
Ref.~\cite{nf2-I} are checked by using an exact RHMC algorithm.
In Section~\ref{firstord} we directly test the first
order hypothesis by using a new set of numerical
simulations performed by keeping 
$a m_q L_s^{y_h}$ fixed with $y_h = 3$. Our conclusions and
perspectives for the continuation of our program will be presented
in Section~\ref{conclusions}.

\begin{figure*}
\includegraphics*[height=.7\columnwidth]{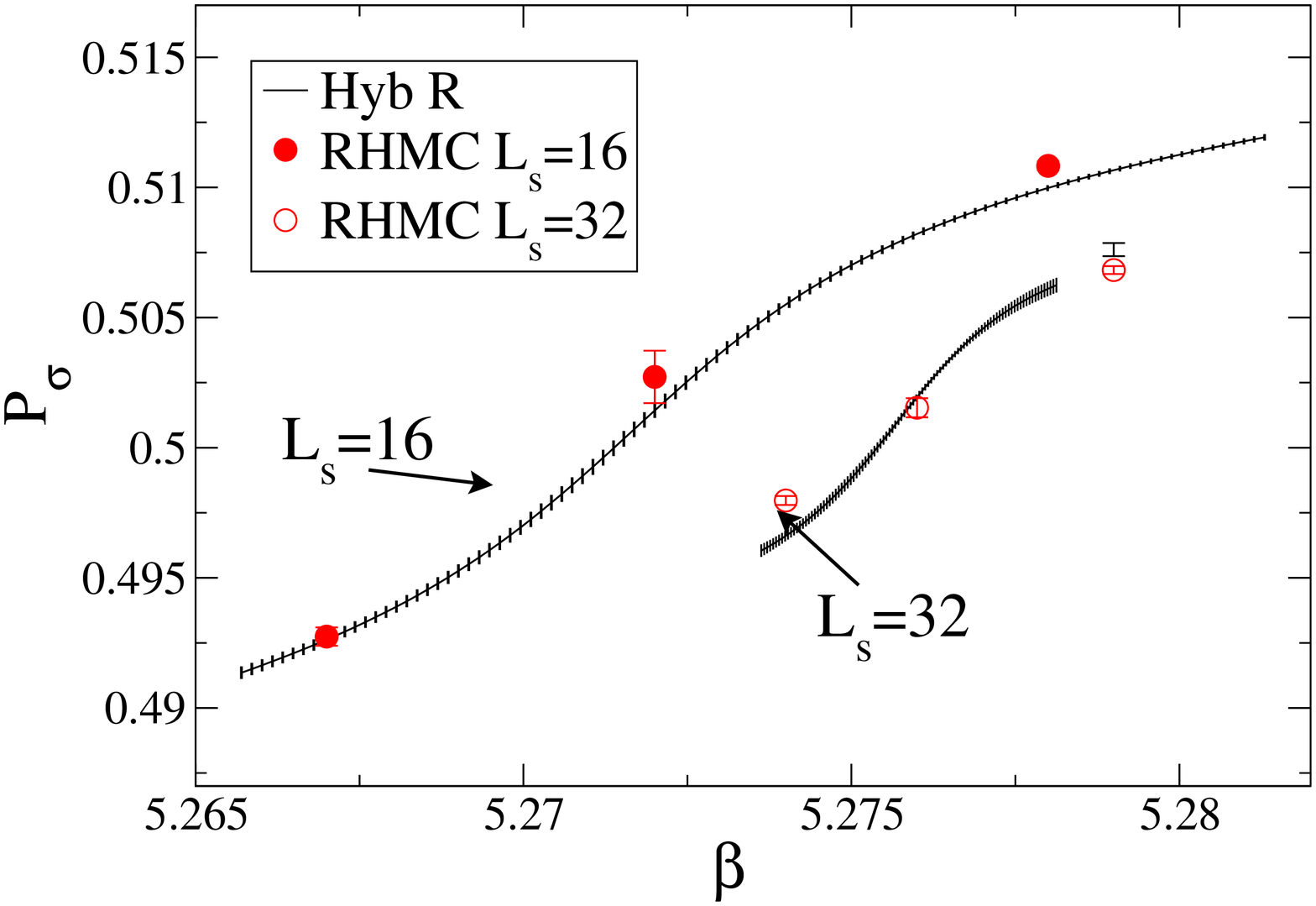}\hfill\includegraphics*[height=.7\columnwidth]{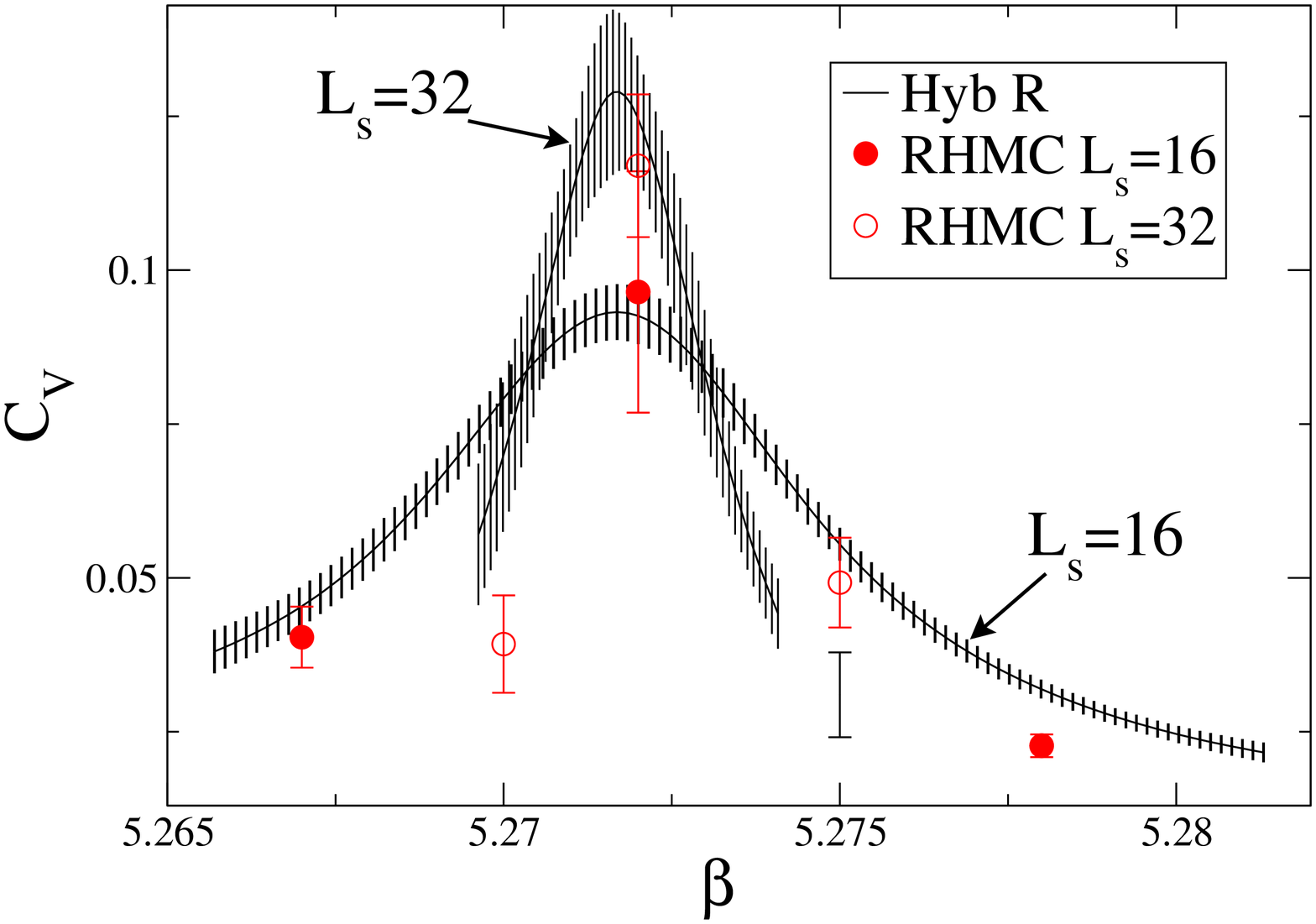}\\
\hfill\includegraphics*[height=.7\columnwidth]{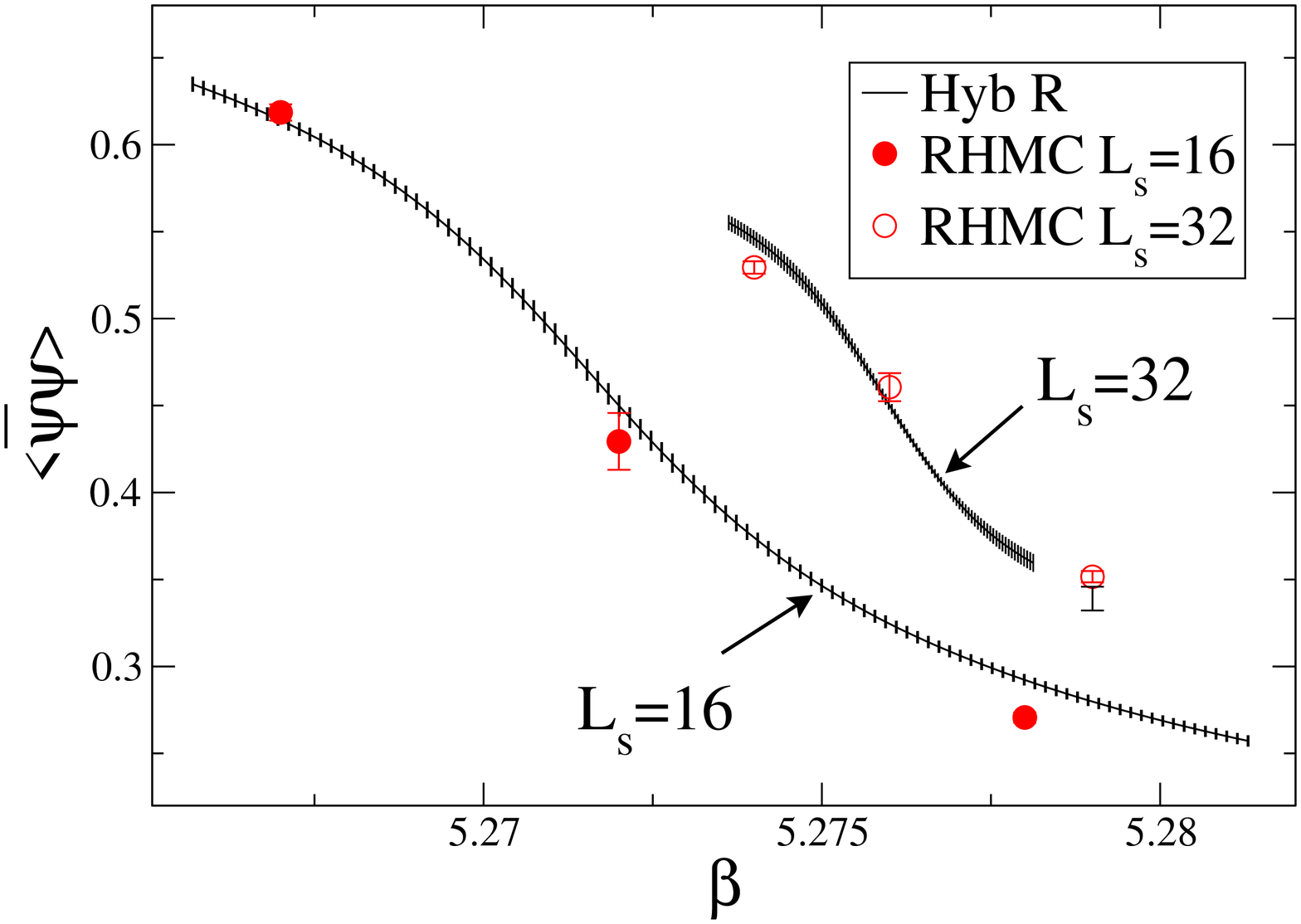}\hspace{.75cm}\includegraphics*[height=.7\columnwidth]{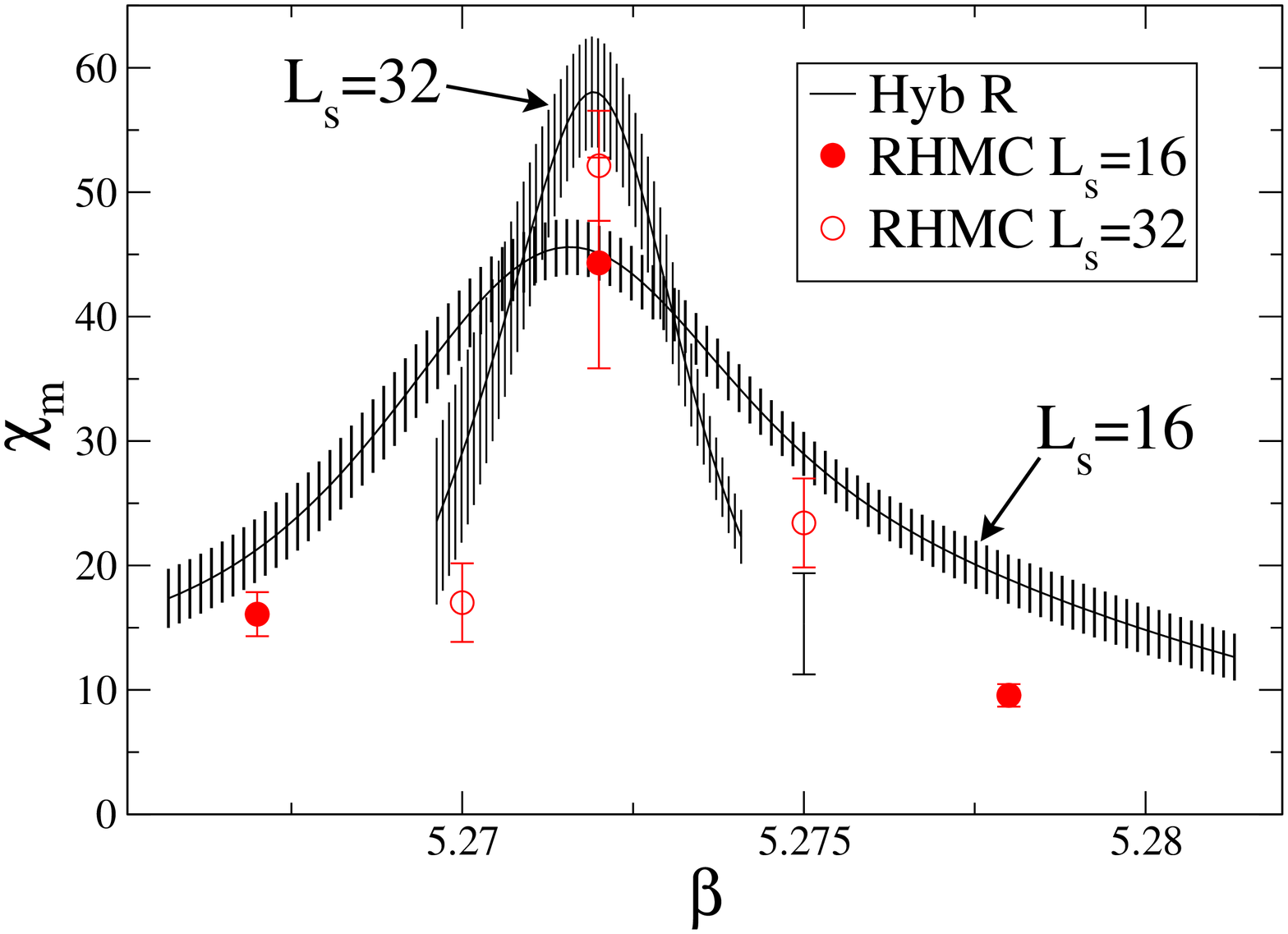}
\caption{Comparison between MC estimates obtained by the exact RHMC algorithm and the Hybrid-R for the average value (left column) and susceptibility (right) of the plaquette (top row) and chiral condensate (bottom) on lattices with $L_s=16,32$ and $m_L=0.01335$. The average values for $L_s=32$ in the left figures have been shifted to the right ($\Delta\beta=0.004$) for sake of visual clarity.\label{RHMCRFIG}}
\end{figure*}

\section{Results obtained with the RHMC algorithm}
\label{rhmc}
Keeping under control the systematic errors introduced by a non-exact
algorithm like the Hybrid-R is a very expensive task in terms of
computer power. In principle one should carry out multiple MC
simulations with smaller and smaller molecular dynamics integration
steps for every value of the simulation parameters. This extrapolation
is in practice unfeasible - especially if one aims at the
investigation of the chiral limit. What can be done is to choose the
integration step size as a function of the simulation parameters, in
our case the lattice quark mass $am_q\equiv m_L$, in such a way that
discretization errors are negligible as compared to the statistical
ones. The right functional form can be determined by a preliminary
study on a representative subset of the parameters or using known
results present in the literature (see e.g.~\cite{jlqcd}). For
standard Kogut-Susskind fermions the simple choice: $\delta t=m_L/4$
is believed to lead to an accuracy of $\approx5\%$ in the
thermodynamic susceptibilities for masses as low as
$m_L\approx0.01$. That was the choice we used in Ref.~\cite{nf2-I} 
for all but the largest volume $L_s=32$ at the smallest mass
$m_L=0.01335$ where, for computational limitations, we took $\delta
t=m_L/2$, which is expected to introduce errors of about $10\%$.

\begin{figure}
\includegraphics*[width=\columnwidth]{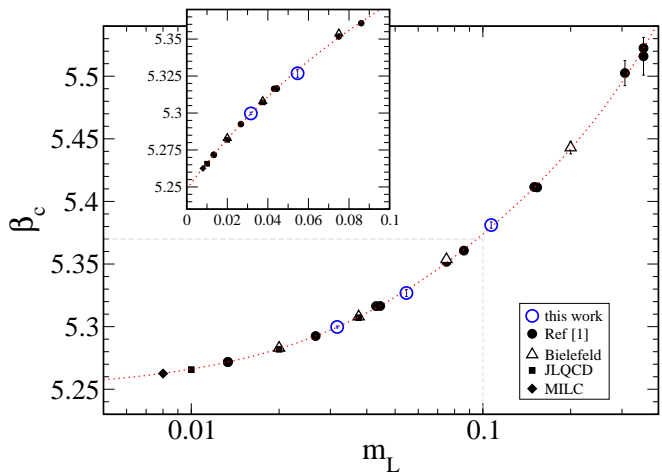}
\caption{Pseudocritical coupling curve in the $(\beta,m_L)$ plane. The
  newly added points are drawn using open circles. To be compared with
  Fig.~3 of Ref.~\cite{nf2-I}.\label{BETAC}}
\end{figure}

Recently the RHMC algorithm~\cite{RHMCALG} has emerged as a convenient
and exact algorithm for staggered fermion simulations. Not only it has
no systematic errors to control (i.e. no estrapolation is needed),
but it also outperforms the R algorithm for 2 flavors in terms of
computer time. This makes the RHMC algorithm the ideal candidate to
put the data of Ref.~\cite{nf2-I} to the test (and to produce the new data).

The subset of simulations used for this comparison was the
one with the smallest lattice quark mass, namely $m_L=0.01335$ with
$L_s=16,32$, where discretization errors are expected to be more
significant and where a larger $\delta t$ was used for the biggest
lattice. We made three different simulations for each value of $L_s$
at couplings just below, just above and at the transition
point. For each value of $\beta$ we accumulated a statistics of order
3k trajectories for $L_s=32$ and of order 15k trajectories for
$L_s=16$. 

The results are shown in Fig.~\ref{RHMCRFIG}. The two most significant
quantities, the average plaquette and the chiral condensate, used in
Ref.~\cite{nf2-I} are displayed together with their respective
susceptibilities for both the lattices. 

A clear agreement at a $1\sigma$ level for all of these quantities is
observed at the transition point. The other points agree within
$2\sigma$ except for the one corresponding to $L_s=16$ above the
transition. The reason for this discrepancy can be traced back to the
limited statistics used for the end tail in this simulation of
Ref.~\cite{nf2-I}, in a coupling region of small importance.
In the scaling region of interest the results obtained in
Ref.~\cite{nf2-I} are proved to be well within the expected errors.

We confidently conclude that the use of the R algorithm in
Ref.~\cite{nf2-I} did not introduce artefacts invalidating the
finite-size scaling analysis.
 
\section{Direct test of the first order hypothesis} 
\label{firstord}

Some hints of a first order transition were observed in
Ref.~\cite{nf2-I}, by use of Eqs.~(\ref{scalcal2})-(\ref{scalord2})
using the dataset generated to
directly check the $O(4)$, $O(2)$ universality classes.
We now repeat the scaling analysis of Ref.~\cite{nf2-I} assuming
weak first order.
We generate a dataset with
a fixed value of $am_qL_s^{y_h}$, with $y_h=3$ as expected for a first
order transition. The most convenient way to proceed is to use the
already available and checked MC simulations at $L_s=32$ and
$m_L=0.01335$. This enables a major saving in computer time.
This also automatically fixes the value of $m_LL_s^3\simeq 437.45$.
Three other sets of simulations were made to construct the dataset for
the first order scaling test: one with $L_s=16$ and $m_L=0.1068$, one with
$L_s=20$ and $m_L=0.054682$ and
one with $L_s=24$ and $m_L=0.03164444$. For each of them ten different
values of $\beta$ spanning the entire critical region were
simulated with a total statistics of about 90k trajectories collected for
each of the lattices. 

The pseudocritical couplings $\beta_c$ of the three new lattices are in
excellent agreement with the pseudocritical curve determined in
Ref.\cite{nf2-I} (see Fig.~\ref{BETAC}). Given such an
agreement, no modification is necessary to Sect.~IVA of Ref.\cite{nf2-I}.

\begin{figure*}
\includegraphics*[height=.7\columnwidth]{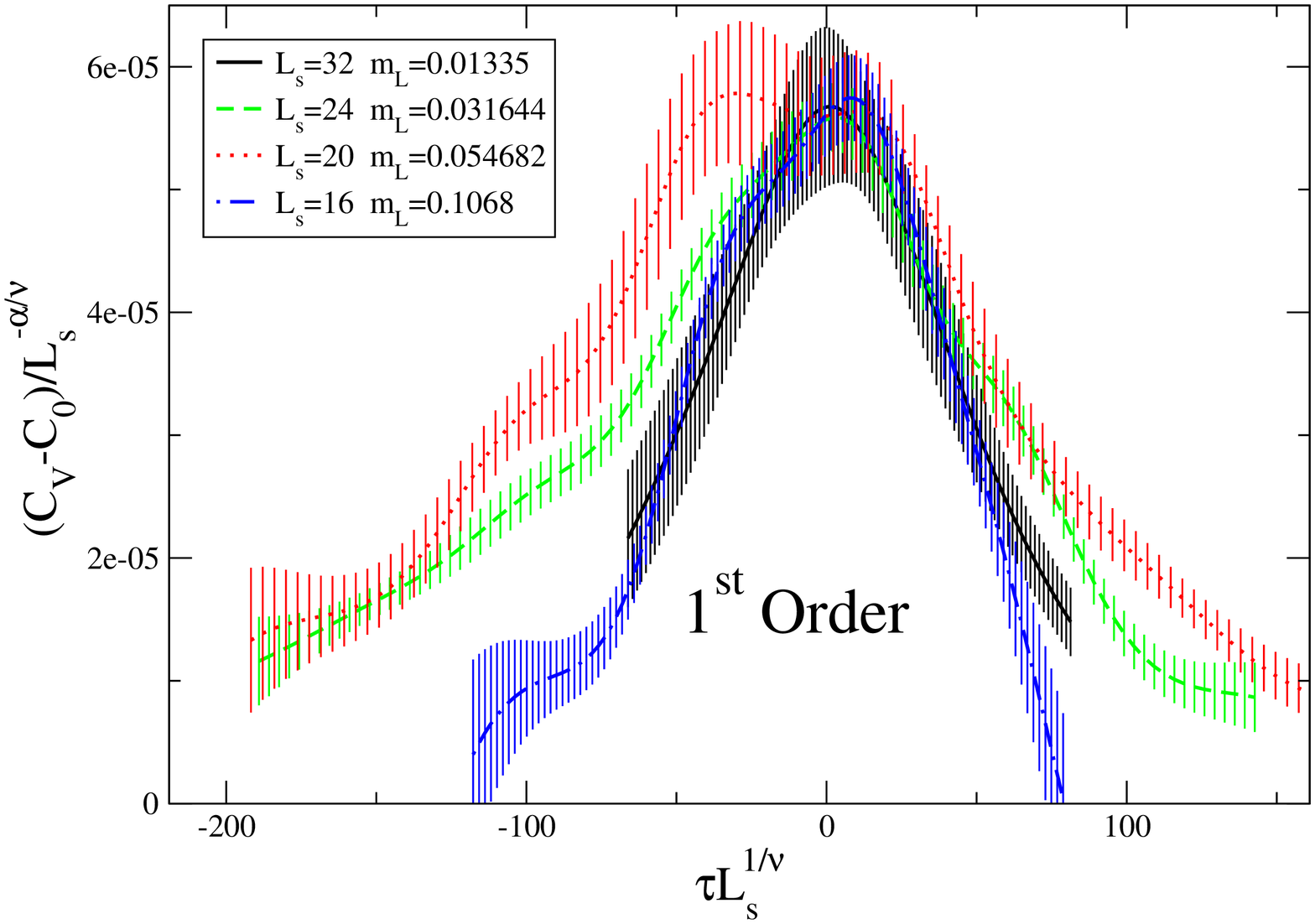}\hfill\includegraphics*[height=.7\columnwidth]{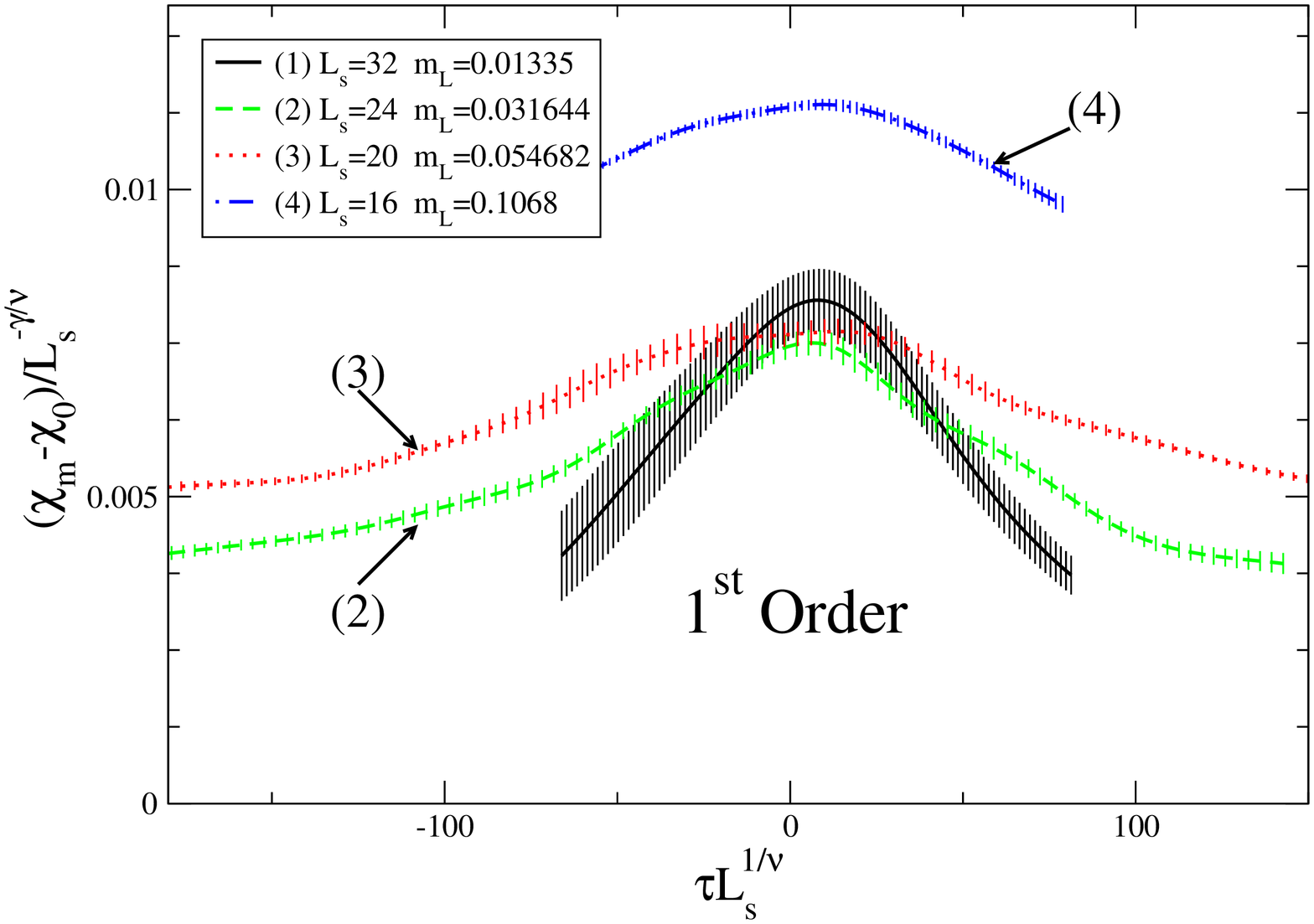}
\caption{Finite size scaling at $m_LL_s^3$ fixed of the subtracted specific heat $C_v-C_0$ and of the chiral condensate susceptibility $\chi_m-\chi_0$.\label{RUN3SCAL}}
\end{figure*}

As in Ref.\cite{nf2-I} the background must be determined in
order to check the scaling. For the case of the specific heat such
background was observed not to depend on $m_L$ and to be almost
linear in $\beta$ in the mass region of relevance for our
purposes. The three new datasets at $m_L=0.03164444$, $m_L=0.054682$
 and $m_L=0.1068$ nicely fit
together with all other data. The background estimate of
Ref.\cite{nf2-I} is only slightly modified by the new data (shifts of
order $0.3\sigma$): $C_0(\beta)= 0.417(51)-0.0695(93)\beta$.  For the
chiral condensate susceptibility $\chi_m$ the background is determined
by a best fit of the peaks of the curves to the function:
$\chi_{peak}(L_s)=\chi_0+kL_s^{\gamma/\nu}$ with $\gamma/\nu=3$,
appropriate to a first order transition (see Eq.~(\ref{scalord})).

The consistency check of the first order finite size scaling is shown
in Fig.~\ref{RUN3SCAL} (the analogous figures for $O(4)$ are Fig.~6 and
17 of Ref.~\cite{nf2-I}). 
A reasonable scaling is observed for the specific heat $C_V$.
As already stated in Ref.\cite{nf2-I}, $C_V$ is independent of any 
prejudice on the symmetry and on the order parameter.
Violations of the scaling Eq.~(\ref{scalord}) are observed for $\chi_m$
at larger values of the masses.
In fact Eq.~(\ref{scalord}) is expected to be valid for the susceptibility 
of the order parameter. At large masses chiral symmetry is badly broken 
and $\langle\bar\psi\psi\rangle$ is possibly not a good order parameter.
In our data, both of the present paper and of Ref.~\cite{nf2-I}, 
Eq.~(\ref{scalord}) seems to be violated for $m_L>0.05$.

\begin{figure*}
\includegraphics*[height=.7\columnwidth]{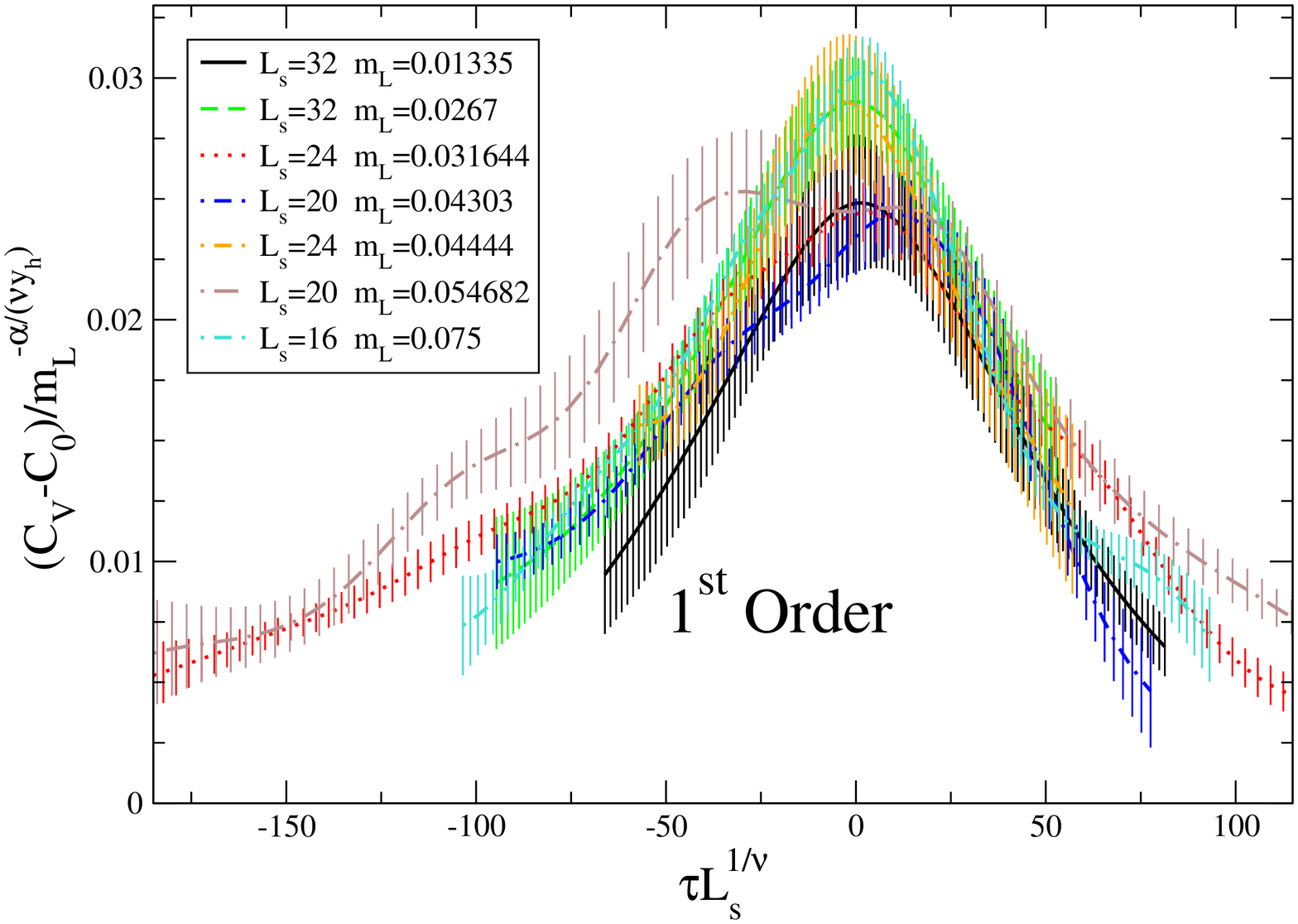}\hfill\includegraphics*[height=.7\columnwidth]{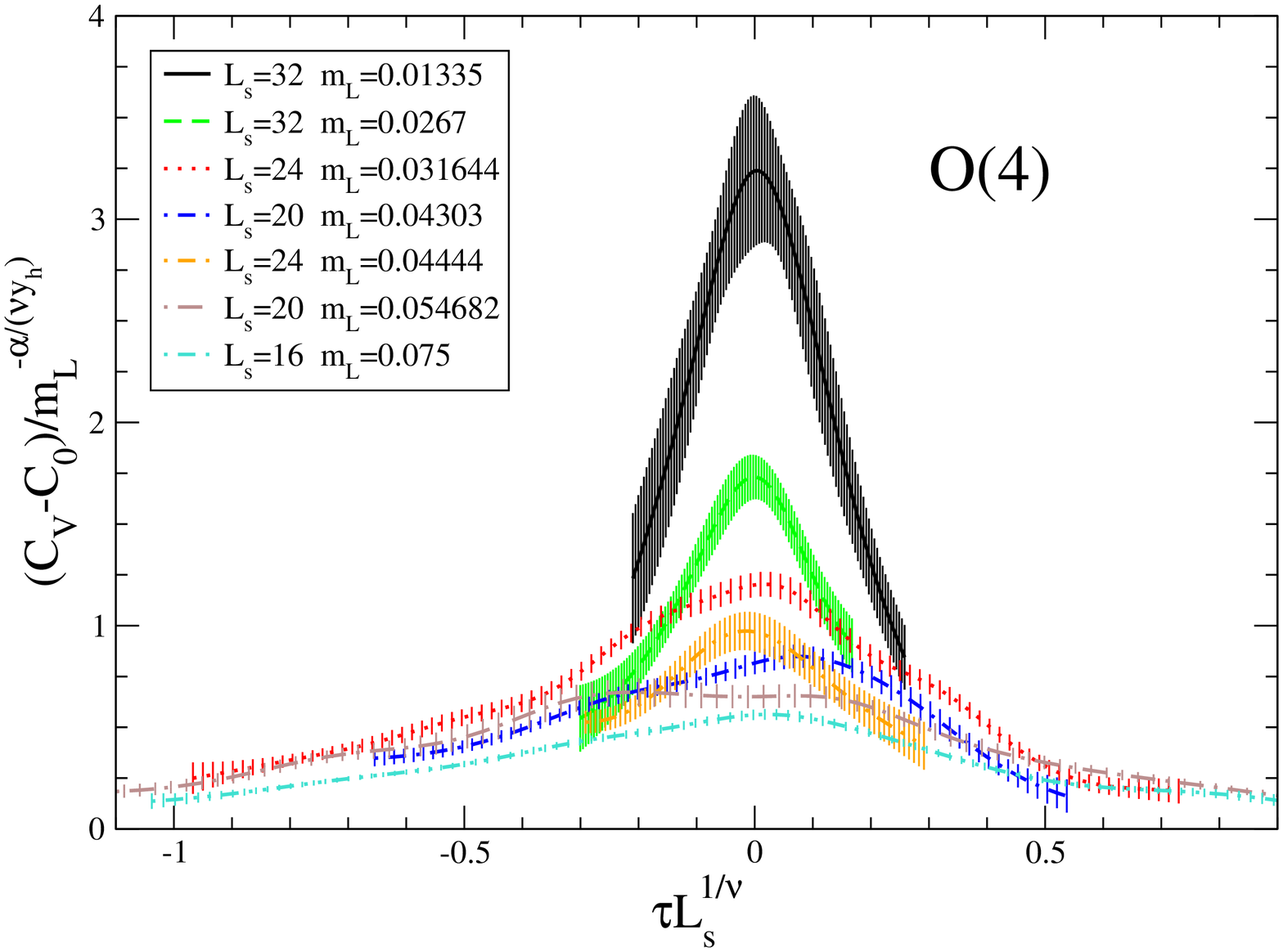}\\
\hfill\includegraphics*[height=.7\columnwidth]{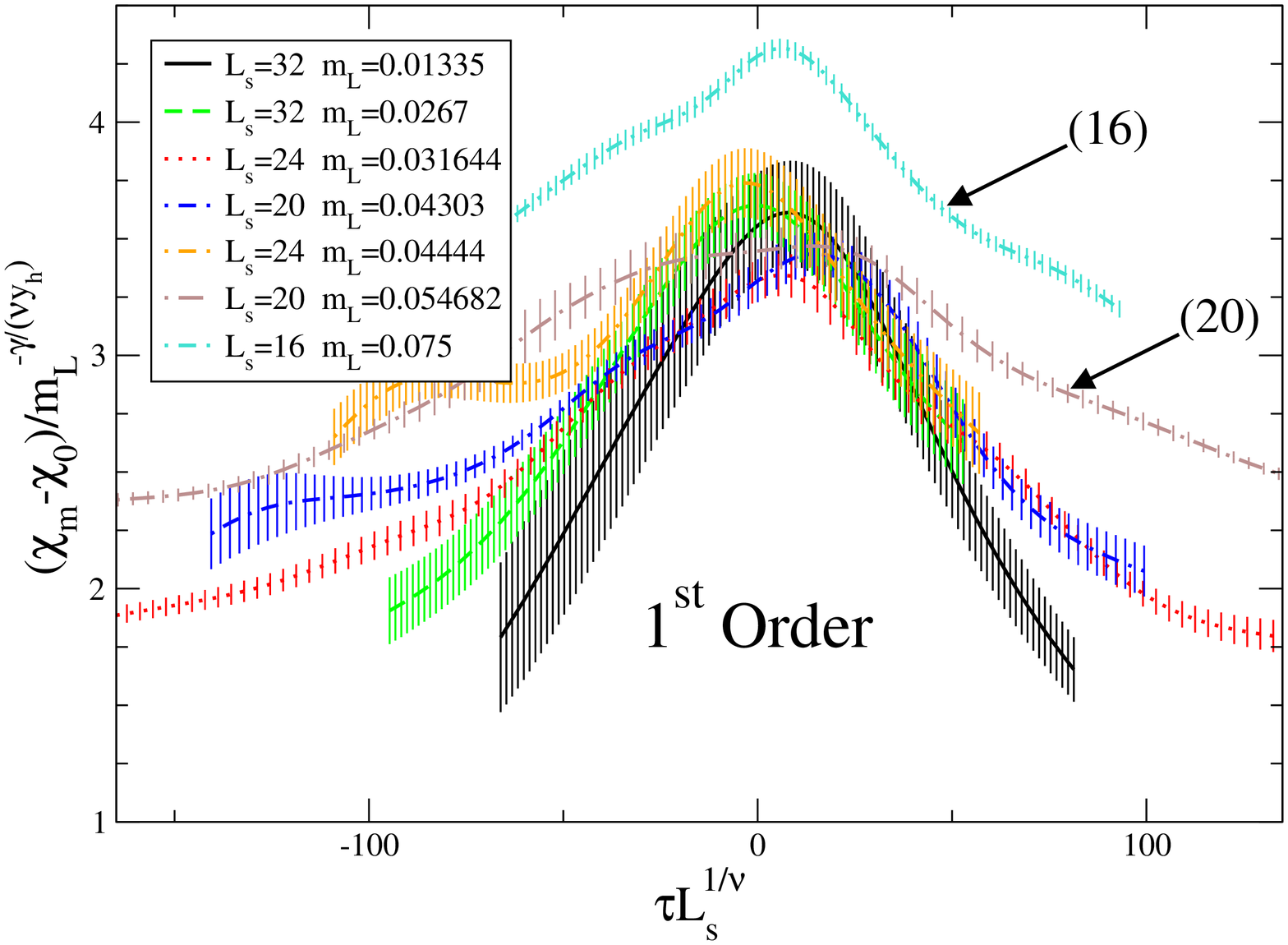}\hspace{.80cm}\includegraphics*[height=.7\columnwidth]{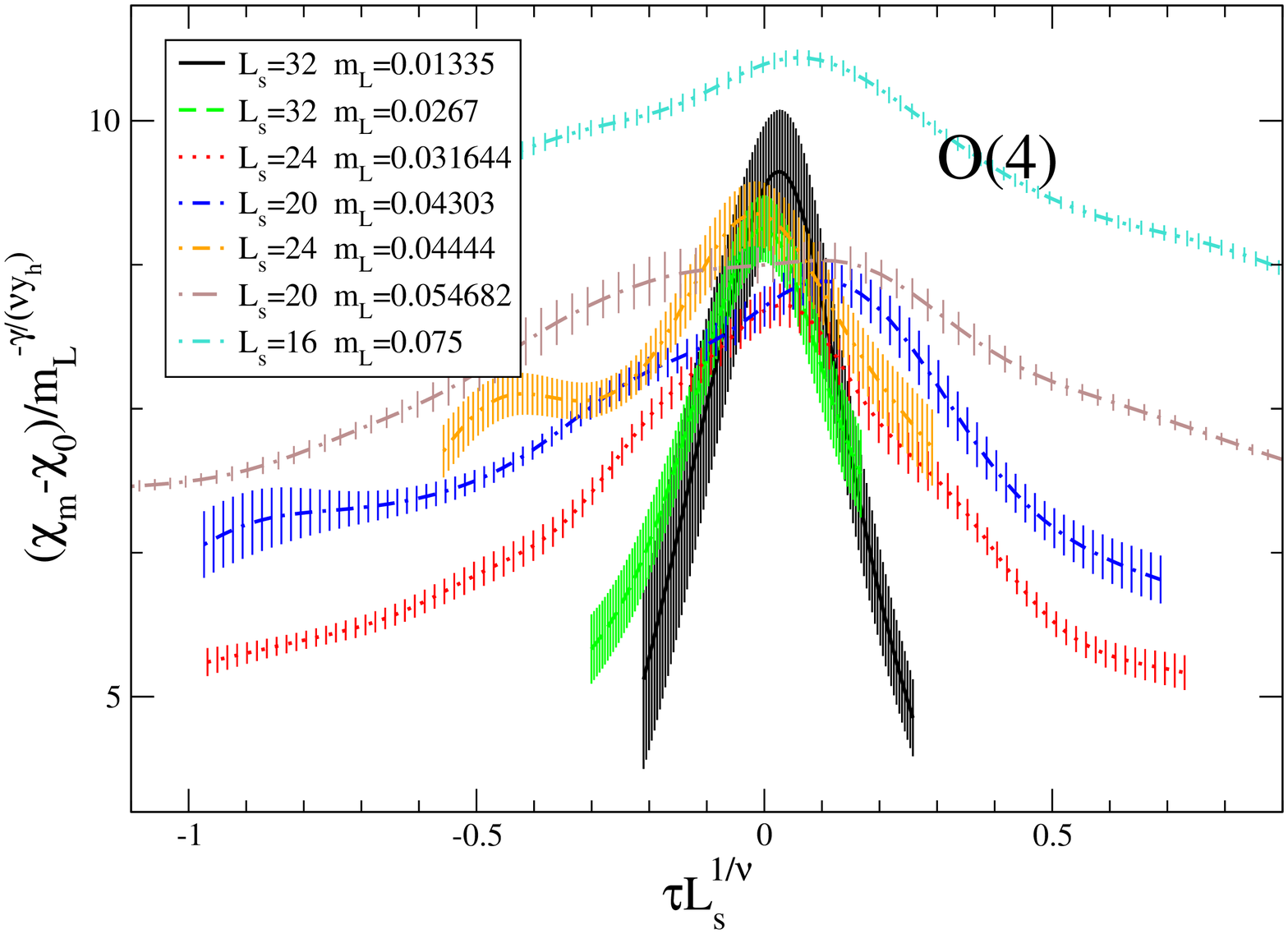}
\caption{Approximate scaling laws corresponding to Eqs.~(\ref{scalcal2})-(\ref{scalord2}). In the fit to the background the curves (16) and (20) with $m_L>0.05$ have been discarded.\label{FO}}
\end{figure*}

\begin{figure}
\includegraphics*[width=\columnwidth]{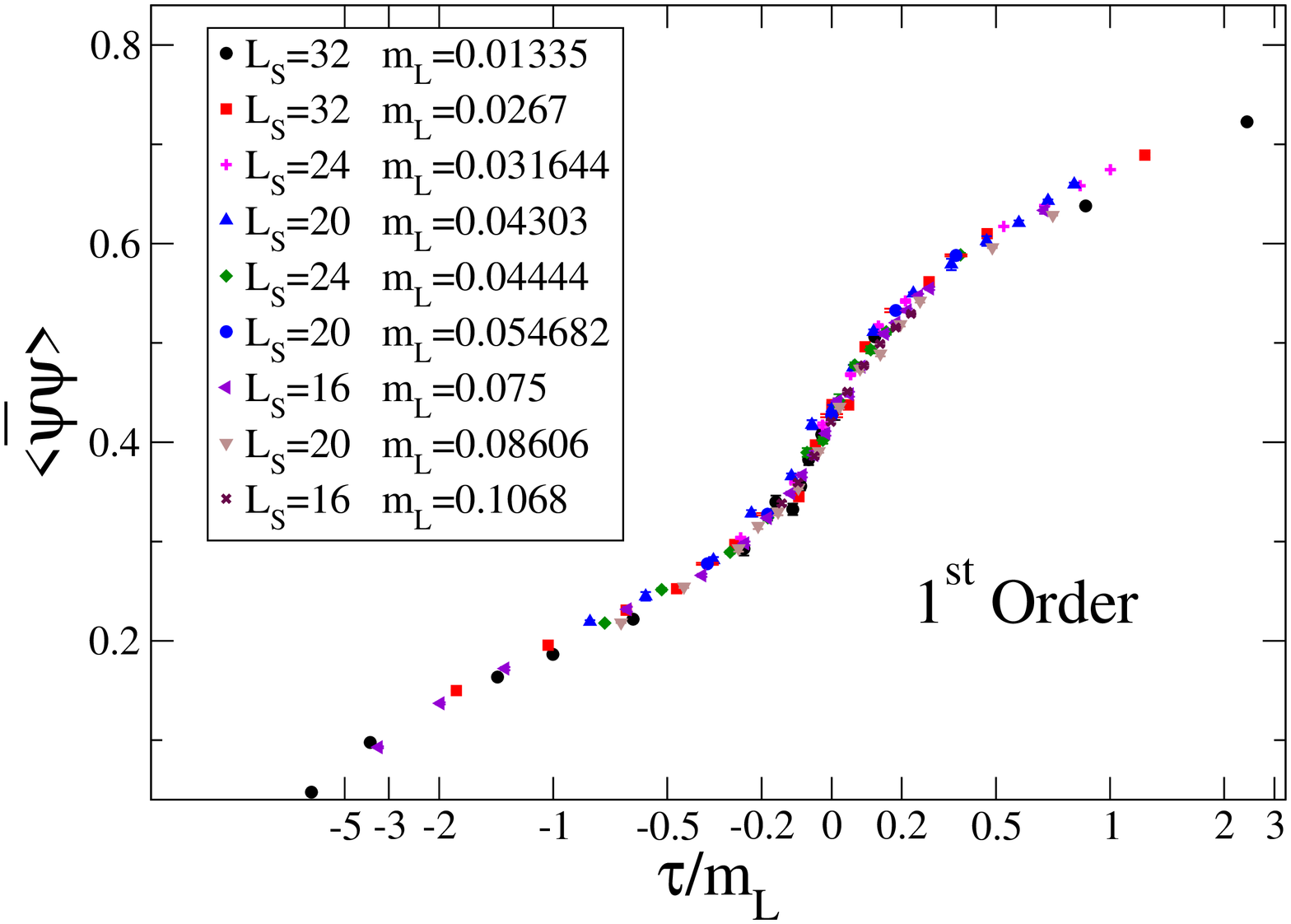}
\caption{Scaling of the subtracted chiral condensate with first order exponents. (x axis has been rescaled with the function $\rm{atan}(x)$ to emphasize the scaling region)\label{EQST}}
\end{figure}

An updated version of the scaling laws Eqs.~(\ref{scalcal2})-(\ref{scalord2}) 
(Fig.~8,~9 and 18 of Ref.~\cite{nf2-I}) with the new data is shown for completeness
in Fig.~\ref{FO}. The scaling
laws Eqs.~(\ref{scalcal2})-(\ref{scalord2}) correspond to taking the
limit $m_LL_s^{y_h}\gg 1$ with $\tau L_s^{1/\nu}$ fixed in
Eq.(\ref{scal2}). In the mass region considered a
reasonable scaling is observed for first order for the specific heat 
(similar considerations as above apply). For the chiral susceptibility
good scaling is observed only for $m_L<0.05$.

\begin{figure*}
\includegraphics*[width=\textwidth]{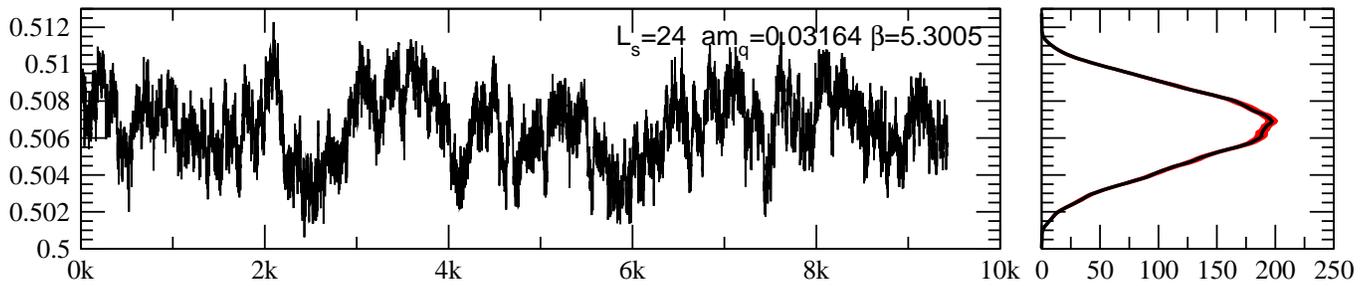}
\caption{MC time history at the pseudo-critical coupling for $L_s=24$,$m_L=0.03164444$. No clear metastabilities are observed.\label{HIST24}}
\end{figure*}

Both in Fig.\ref{RUN3SCAL} and in Fig.~\ref{FO}, the background for the chiral 
susceptibility has been obtained by the fit to the curves with $m_L<0.05$.
The fact that only in a neighboorhood of $m_L=0$ the scaling is 
expected is well known -- but the actual mass range is not.
The fitting procedure described allows us to identify a scaling 
window where the whole curve scaling can be verified. 

The scaling of the chiral condensate (magnetic equation of state) can
also be checked. The expectation for first order is:
\beq
\langle\bar\psi\psi\rangle = f(\tau m_L^{-1})\label{MAGEQ}
\eeq
In Ref.~\cite{nf2-I} we found that if a subtraction is allowed on the
left-hand side of Eq.~(\ref{MAGEQ}) then a good scaling is found with the
first order exponent. The subtraction was found to be numerically
equal to the value of the condensate on the zero gauge field background 
(perturbative value). This,
in the region explored, is equivalent to impose that all the
curves coincide at the pseudocritical coupling. 

The updated figure for the first order scaling is shown in
Fig.~\ref{EQST} (see Fig.~14 of Ref.\cite{nf2-I}). The
scaling is very good within the statistical errors. The three
new curves have been inserted and have been found to scale nicely.

As a last check we looked for metastabilities in the MC histories of
the new lattices. We found no metastable states. The history
corresponding to the pseudo-critical coupling of the $L_s=24$ lattice
is shown in Fig.~\ref{HIST24}. No clear double-peak structure is found. If
the transition is first order, increasing the lattice size, one should
eventually see the metastabilities. At present, no clear sign of these
states are found in the numerical data.

\section{Conclusions}
\label{conclusions}
The determination of the order of the chiral transition for $N_f=2$ QCD
has turned out to be a very challenging problem because of the huge computational 
resources required.

In Ref.~\cite{nf2-I} we proposed a novel method to deal with the double scaling
Eq.~(\ref{scalcal}),(\ref{scalord}).
We applied such method to test the $O(4)$, $O(2)$
critical behavior -- expected for a second order transition on the basis of theoretical
speculations. The conclusion was that our numerical data was not compatible with 
$O(4)$, $O(2)$ and we found hints of a first order transition.

In the present work we have completed our previous analysis in two different respects.

In Sect.~\ref{firstord}, we have performed a direct test of first order scaling, 
using new MC data with parameters, with $m_LL_s^3=437.45$, 
$L_s=16,20,24,32$.
These lattices show a good scaling for the specific heat and
for the chiral susceptibility at masses $m_L<0.05$ (see Fig.~\ref{RUN3SCAL}).  
At larger masses scaling of the chiral susceptibility is broken, presumably because of strong breaking of 
the chiral symmetry.

For the sake of completeness we have also updated the scaling pictures of 
Ref.~\cite{nf2-I} corresponding to Eqs.~(\ref{scalcal2}),(\ref{scalord2}) (see Fig.~\ref{FO}).

The good scaling of the chiral condensate with first order pseudo-critical 
exponents, observed in Ref.~\cite{nf2-I}, is also in complete agreement with the new MC data
produced for this work (see Fig.~\ref{EQST}).

In Sect.~\ref{rhmc}, we have checked some of the systematic errors present in the previous analysis
of Ref.~\cite{nf2-I}, by comparing a representative subset of the old MC data generated using 
the non-exact R-algorithm with a new one obtained with the exact RHMC algorithm.
The direct comparison has shown no significant deviations between the two datasets at the critical 
coupling -- thus validating the result presented in Ref.~\cite{nf2-I}.

Taking all the evidence together, the first order scaling is clearly preferred over the 
second order $O(4)$ behavior. 
We can say that $O(4)$, $O(2)$ are excluded and first order is consistent with data, modulo possible 
effects due to the discretization. Again we think that ultraviolect effects should 
be irrelevant with respect to the large volume behavior. However
the use of finer lattices and new simulation algorithms
to approach the chiral limit will possibly clarify this issue.

\section{Acknowledgments}
We acknowledge the technical staff of the APENEXT centre of INFN (Rome) where most
of the simulations were performed.

The work of C.P. has been supported in part by contract DE-AC02-98CH1-886
with the U.S. Department of Energy.

\end{document}